\documentclass{jkas}

\def\year{2018} 
\def\month{August} 
\def\volume{00} 
\def\issue{0} 
\def\beginpage{1} 
\def\endpage{0} 
\def\received{XX YY, 2018} 
\def\accepted{XX YY, 2018} 
\date{Received \received; accepted \accepted}

\usepackage{color}

\title{
Measuring Timing Properties of PSR B0540$-$69
}

\author{Minjun Kim and Hongjun An}

\affil{Department of Astronomy and Space Science, Chungbuk National 
University, Cheongju, 28644, Republic of
Korea; \email{iehatro@chungbuk.ac.kr, hjan@chungbuk.ac.kr}}

\def\corrauthor{Hongjun An

}

\def\runningauthor{

}

\def\runningtitle{
Timing Properties of PSR B0540$-$69
}

\def\keywords{pulsars: general -- pulsars: individual (PSR B0540$-$69)

}

\def\abstracttext{
We report on the timing properties of the `Crab twin' pulsar PSR~B0540$-$69
measured with X-ray data taken with the \textit{Swift} telescope over a period
of 1100\,days. The braking index of the pulsar was estimated to be
$n = 0.03\pm0.013$ in a previous study performed in 2015 with 500-day {\it Swift} data.
This small value of $n$ is unusual for pulsars, and a comparison
with an old estimate of $n\approx2.1$ for the same target determined
$\sim$10 years ago suggests a dramatic change in the braking index.
To confirm the small value and therefore the large change of $n$,
we used 1100-day {\it Swift} observations including the data used in the earlier determination of
$n=0.03$. In this study we find that the braking index of PSR~B0540$-$69 is $n=0.163\pm0.001$,
somewhat larger than $0.03$. Since the measured value of $n$ is still much
smaller than $2.1$, we can confirm the dramatic change in the braking index for
this pulsar.
}

\begin{document}
\jkashead 


\section{Introduction\label{sec:intro}}

Neutron stars are the left-over cores of massive stars after a supernova explosion.
They are composed of extremely dense matter and have very strong magnetic fields ($\sim10^{12}$\,G).
They are usually detected as pulsating (rotating) sources in the radio to 
gamma-ray band \citep{Ostriker}.
Their spin frequency slowly decreases and timing properties in this spinning
down process can be characterized by measuring the spin frequency $\nu$ and its
time derivatives $\dot\nu$ and $\ddot \nu$ which are combined to derive the braking
index $n=\ddot\nu\nu/\dot\nu^{2}$ \citep{Manchester}.

Measuring the braking index for pulsars is difficult because the measurement
requires an estimation of the second derivative which is in general very small. 
In addition, the rotation of neutron stars is known to be irregular due to `glitches', 
sudden changes in the spin-down rate and to `timing noise', long-term non-period
modulations. Thus pulsars' timing properties can be characterized only by observing
the pulsars for a long time. Up to now braking indices have been measured for only about 10 
of the 2600 detected pulsars (Table.~\ref{tab:brakingindecies}) and are typically less than 3 \citep{Lyne2015}.

The braking index of a pulsar can be used to understand the energy-loss mechanisms of the pulsar. 
For ideal magnetic dipole radiation, the braking index is expected to be around 3 \citep{Lyne2015},
but it can differ due to other energy loss mechanisms (e.g., wind and gravitational wave) 
or to changes in the mechanical properties. $n$ can be $\sim$1 if energy loss of a pulsar
is dominated by the wind or 5 if gravitational-wave radiation is dominant \citep{Kou, Araujo}.
Effects due to changes in the mechanical properties on $n$ are very complex.
Several of these effects may operate at the same time, and by measuring $n$, one can infer
which effect is dominant. The evolution of a pulsar can be studied because the
changes on $n$ can imply any effect of the aforementioned mechanisms.

Until now, a wide range of values for $n$ has been measured.
Although the measured braking indices for various pulsars can be between
$-1.2$ \citep[PSR~J0537$-$6910, perhaps contaminated by frequent glitches;][]{Antonopoulou} or
0.9 \citep[PSR~J1734$-$3333;][]{espinoza}
and 3.15 \citep[PSR~J1640$-$4631;][]{Archibald2016}, most cases are only within $2 \le n \le 3$ \citep{Lyne2015}.
Furthermore, $n$ has been observed to change in some pulsars;
the braking index of PSR~J1119$-$6127 changed from 2.91 to 2.684
when a glitch occurred \citep{antonopoulou2015} and that of PSR~J1846$-$0258 changed from 2.65 to 2.19
after an outburst \citep{Archibald}.
While it is not yet clearly understood how various pulsars have such values of
$n$ and how they change, some theories have been developed: accretion from a hypothetical
fall-back disk \citep{Menou}, super-fluid decoupling \citep[][]{Antonopoulou},
magnetic-field evolution \citep[][]{Gourgouliatos, blandford}, and
an effective change in the moment of inertia \citep{sedrakian}.
Because there are not many pulsars with a measured braking index, understanding 
the value and its change on a firm theoretical ground is still lacking. 
So it is crucial to measure $n$ and its changes for more sources to 
find extremes to constrain the models.

\begin{table*}[ht]
\centering
\caption{List of pulsars with a measured braking index \label{tab:brakingindecies}}
\centering
\scriptsize{
\begin{tabular}{lccccc}
\toprule
Pulsar & $\nu$ (Hz) & $\dot\nu$ $(10^{-10}s^{-2})$ & Characteristic Age(kyr) & $n$ & Ref. \\
\midrule
J0534$+$2200 & 29.946923 & $-$3.77535 & 1.26 & 2.342(1) & \citet{Lyne2015}\\
J0537$-$6910 & 62 & $-$1.992 & 4.93 & $-$1.22(4) & \citet{Antonopoulou}\\
J0835$-$4510 & 11.200 & $-$0.15375 & 11.3 & 1.7(2) & \citet{espinoza2017}\\
J1119$-$6127 & 2.4512027814 & $-$0.2415507 & 1.6 & 2.91(5)$\sim$2.684(2) & \citet{antonopoulou2015}\\
J1208$-$6238 & 2.26968010518 & $-$0.16842733 & 2.67 & 2.598(1) & \citet{clark}\\
J1513$-$5908 & 6.611515243850 & $-$0.6694371307 & 1.56 & 2.832(3) & \citet{livingstone2011b}\\
J1640$-$4631 & 4.843410287 & $-$0.2280830 & 3.4 & 3.15(3) & \citet{Archibald2016}\\
J1734$-$3333 & 0.855182765 & $-$0.0166702 & 8.13 & 0.9(2) & \citet{espinoza}\\
J1833$-$1034 & 16.15935711336 & $-$0.52751130 & 4.85 & 1.857(1) & \citet{roy}\\
J1846$-$0258 & 3.0621185502 & $-$0.6664350 & 0.73 & 2.65(1)$\sim$2.16(1) &  \citet{Archibald}\\
\bottomrule
\end{tabular}}
\end{table*}

PSR~B0540$-$69 is a young and bright 50-ms pulsar in the Large Magellanic Cloud
 \citep{Seward}.
The pulsar is surrounded by a pulsar wind nebula \citep[PWN, $R_{\rm{pwn}}\approx 4''$;][]{Kaaret}
which has a similar morphology \citep[torus and jet;][]{Campana, Mignani2012} to that
of the Crab nebula \citep{mori,madsen}.
The rotation properties are well measured to the second derivative with a characteristic age
$\tau _c = \nu /2\dot\nu$ of $\sim 1700$\,years and a braking index of $n\approx 2.1$ \citep[e.g.,][]{Livingstone}.
The most interesting property of this pulsar is a large change in the braking index which
was measured to be $n=2.123\pm0.012$ between 1999 and 2011 with {\it RXTE} in $2-60$\,keV band \citep{Ferdman},
but a recent study reported $n=0.031\pm0.013$ using the 500-day {\it Swift} observations taken
between 2015 and 2016 \citep{Marshall2016}. Such a large change and a small value of $n$ 
have not been seen in any other pulsars. The braking index of $n\approx 2.1$ before 2010
seems to be reliable because several works have obtained similar values
($n\approx 2.1$) by independently analyzing the long ($>$10\,years) {\it RXTE}
and/or {\it BeppoSAX} ($2-30$\,keV) observations 
\citep[][]{Cusumano, Livingstone, Ferdman,Zhang}.
However, $n=0.03$ derived from 500-day {\it Swift} data in \citet{Marshall2016} has not been carefully examined. Thus, a careful reinvestigation is needed.

In this work, we reanalyzed the 500-day {\it Swift} data used in \citet{Marshall2016} to confirm
their timing solution. We then added recent 600-day {\it Swift} observations to extend
the baseline and to refine the braking-index measurement.
We describe the observations in Section~\ref{sec:observation} and present the timing analysis and
the results in Section~\ref{sec:timing}. We then discuss and conclude in Section~\ref{sec:discussion}.

\section{Observations and data reduction\label{sec:observation}}

We used X-ray data of PSR B0540$-$69 observed by the {\it Swift} X-ray telescope \citep[][]{Gehrels}
between 2015 Feb. 17 to 2018 Feb. 11. Note that the baseline of the observations
we used is about twice that used in \citet{Marshall2016}.
In this period, 62 {\it Swift} observations with a typical exposure of $\sim$2\,ks were performed
with the window timing mode (a time resolution of 1.7\,ms)
to facilitate measurements of the short 50-ms period of the pulsar.
We processed the observational data using the standard {\it Swift} pipeline incorporated in
HEASOFT~6.20 along with the latest CALDB calibration files.
We selected events using a $R=20''$ circular region centered at the source position
in the 0.7--7\,keV band. The photon arrival times are then corrected to Solar System's barycenter with 
the source position $\alpha=05^h 40^ m11.202^s$ and $\delta=-69^\circ 19'54.17''$ \citep{Mignani2010}
and the JPL~DE200 ephemeris. Because of the short 94-min orbital period of the {\it Swift} satellite,
long observations suffer from Earth occultation and have observational gaps (occultation).
To remove the artifacts of the occultation (e.g., beats),
we split an observation into continuous segments when the observation is longer than
the {\it Swift} orbital period.
After pre-calibration and selection, there were $208-1369$ photons per segment.

\section{Timing Analysis\label{sec:timing}}

Young pulsars like PSR B0540$-$69 tend to be more active (e.g., more glitches and timing noise);
therefore, their timing properties seem to be more irregular \citep[e.g.,][]{Dib}.
Indeed, \citet{Livingstone} and \citet{Ferdman} found some glitches and
relatively large timing noise. The latter is a particular concern in the results
of \citet{Marshall2016} because they did not find any timing noise. 
Non-detection of timing noise in their work implies that the pulsar timing noise was 
significantly smaller in 2015 than in 2010.
Indeed, \citet{Marshall2016} noted that there could be small timing
noise but it would not change their results significantly.
Alternatively, it is also possible that there was significant timing
noise in 2015--2016, but the 500-day observations
could not distinguish between the real timing behavior (timing solution)
and the long-term timing noise because the baseline ($\sim$500\,days)
was too short.\footnote{Usually ``timing solution'' means the real timing behavior and the long-term
noise together. Here, however, we separate them for clarity.}
In this case, their timing solution would have been biased due to the noise.
The characteristic time scale of the timing noise in PSR B0540$-$69 was
$\sim$700\,days \citep[e.g.,][]{Ferdman}, and hence, if this trend persisted, 
timing noise can be characterized only by using data with 
a $>700$-day baseline.
Thus, the timing behavior of PSR B0540$-$69 described by the {\it Swift} observations
need to be re-investigated carefully.

There are several ways to determine a pulsar's timing solution. Here we used two
methods: direct period measurements for the individual time series and
phase connection. The former uses the $Z^2_1$-test \citep{jager} to measure 
the periodicity in individual observations. 
This is a robust method for a period detection to avoid any minor irregularity but is not
sensitive to describe the timing properties.
The latter, the phase connection, is more sensitive to
timing noise because it uses both the
periodicity and the pulse shape, but requires that the
observations should be inspected with specific time intervals.
The observation data that we use satisfied the
time-interval requirement for the phase connection method; therefore, we used the more
sensitive technique which enabled us to measure $\ddot \nu$ and the timing noise. 
Aforementioned two methods may result in different solutions when the baseline is long and the pulsar
exhibits a timing anomaly during the observations \citep[e.g.,][]{An2013}.

\subsection{Direct frequency measurements\label{Fourier}}

\begin{figure}
\centering
\includegraphics[width=80mm]{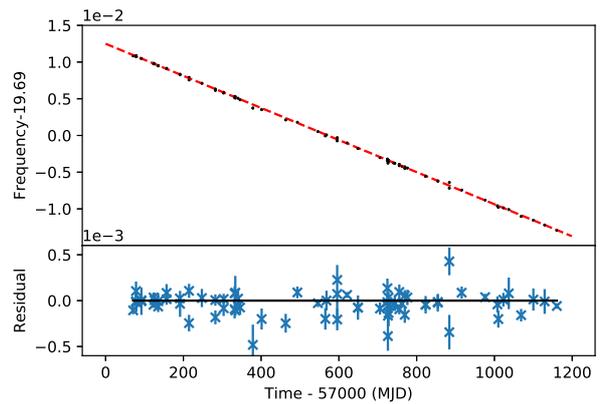}
\caption{Results of the direct frequency measurement using the $Z^2_1$-test.
Top: Frequencies measured for each continuous data segment (black data points) and a linear fit
(red solid line) to measure the frequency derivative. Bottom: residuals after fitting out the linear function.
\label{fig:ftresult}}
\end{figure}

We first used the more robust but less sensitive method, the $Z^2_1$-test \citep[][]{buccheri} for a sanity check,
which basically computes the $Z^2_1$ statistic value in the folded light curve at assumed periods.
The $Z^2_1$ value will be large if there is periodicity at some frequency, so one can find the
best $\nu$ that maximizes the $Z^2_1$ value by scanning $\nu$ near the known value ($\nu\approx 19.696$\,Hz for PSR B0540$-$69).
Uncertainties for $\nu$ measurements were obtained by running simulations.
We did this for each data segment and show the best frequencies that we determined in Figure~\ref{fig:ftresult}.
It is clear from the figure that
the frequency changes with time; hence, we measured the first derivative ($\dot\nu$)
by fitting the frequency trend with a linear function.
The fit was acceptable with $\chi_{\rm min}^2/dof=46.3/71$, and the best-fit frequency and
the derivative are $\nu=19.69633(1)$\,Hz and $\dot\nu=-2.533(3) \times 10^ {-10}\ \rm s^{-2}$ at the epoch of MJD $57281.24147641701$, respectively.
1-$\sigma$ uncertainties are estimated by scanning $\nu$ and $\dot \nu$ to find a range for the parameters
in which $\chi^2$ is less than $\chi^2_{\rm min}+2.3$.
We also fitted the trend with a quadratic function to measure the second derivative 
($\ddot \nu$) but the uncertainty for the quadratic term is large, so we only derived a 90\% 
upper limit of $\ddot \nu < \ddot \nu_{\rm best} + 1.28\Delta \ddot \nu = 2.9\times 10^{-20}\ \rm s^{-3}$, assuming a normal distribution.
Because we cannot measure $\ddot \nu$ with this method, we tried a more sensitive method described below.

\subsection{Phase semi-coherent method\label{coherent}}

\begin{figure}
\centering
\includegraphics[width=80mm]{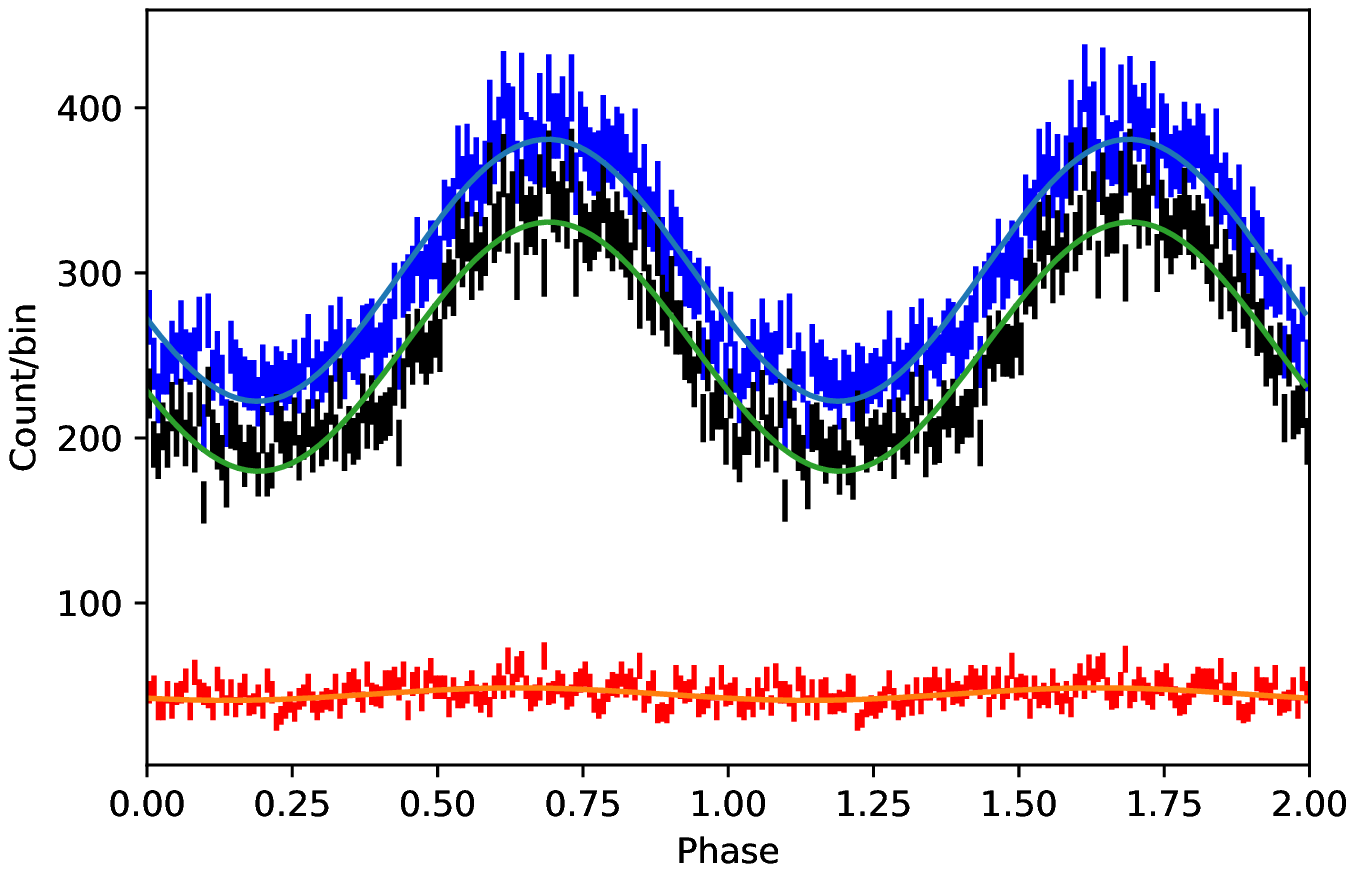} \\
\includegraphics[width=80mm]{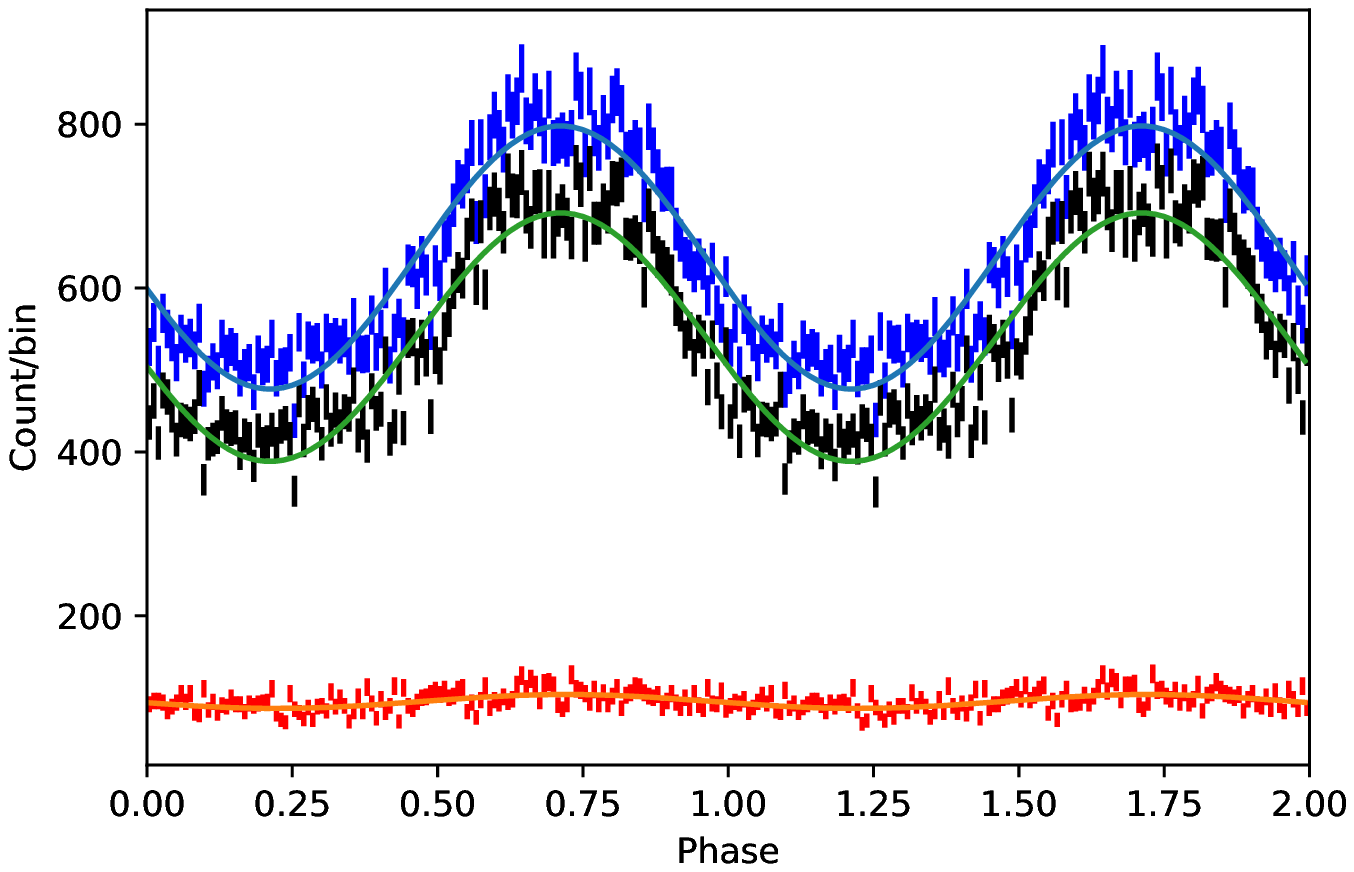}
\caption{0.5--7-keV pulse profiles of PSR B0540$-$69. {\it Top}: the pulse profiles
in the first $\sim$500-day data produced with the known timing solution of \citet{Marshall2016}.
{\it Bottom}: The pulse profile made with the 1100-day data and the refined solution. Blue and black points are pulse profiles of the source before and
after background subtraction, respectively. 
Best-fit sine functions are also shown in lines, and the background profile is shown
in red. We used 128 bins for the profiles and showed two cycles for clarity. The profiles for 500-day and 1100-day data look similar, having pulsed fraction of 0.295(6) and 0.280(4).
\label{fig:marshallpf}}
\end{figure}

We applied the phase-connection method (so-called phase semi-coherent analysis) 
to measure the period second derivative.
The Phase semi-coherent method \citep[e.g., see][]{Livingstone} starts from
an existing timing solution for the first data set (a part of the whole data) to be analyzed
and gradually improves the solution by adding more data. With the first data set and
the timing solution for the data, a rotation phase for each detected photon is
computed using $\phi(t) = \phi_0 + \nu t + \frac{1}{2}\dot \nu t^2 + \frac{1}{6}\ddot \nu t^3$,
where $\phi(t)$ is the rotation phase of the photon detected at time $t$; $\phi_0$ is the
reference phase, and $\nu$ and its derivatives are the timing solution. An initial pulse
profile is generated by constructing the phase histogram (pulse profile, see Fig.~\ref{fig:marshallpf} bottom).
Using the initial timing solution, we construct the pulse profile for the next observation and compare with data.
Although the two pulse profiles can be compared directly, it is
better to use a continuous function to represent the pulse profiles
to reduce the effects of statistical fluctuations.
Thus, we model the profiles with a sine function (Fig.~\ref{fig:marshallpf}) and check to 
see if the profiles align (no relative shift, i.e., phase-connect).
If the phase shift of the new profile with respect to the previous
one is less than 1 cycle ($\Delta \phi<1$), we keep including the next data.
Note that we were not seriously concerned for the fit between the profile and the
sinusoidal function because we compare the relative shift only. As we keep doing
this, the phase shift with respect to the initial profile
may get larger (slow drift with time) and eventually become greater than 1 and thus,
we are not able to phase-connect the new profile to the initial one if the initial
timing solution is not perfect. 
If this occurs, we update the timing solution by varying $\nu$, $\dot \nu$, and 
$\ddot \nu$ to phase-connect all the data up to the point just before the disconnection.
We repeat this procedure until the last observation.
In these process, it may not be possible to phase-connect all the data perfectly (i.e., no relative shift)
with only three timing parameters $\nu$, $\dot \nu$, and $\ddot \nu$, and there may
be small ($\ll1$) relative phase shifts between observations which cannot be modeled
with the polynomial with degree 3. 
These residual phase shifts are usually attributed to timing noise.

\begin{figure}
\centering
\includegraphics[width=80mm]{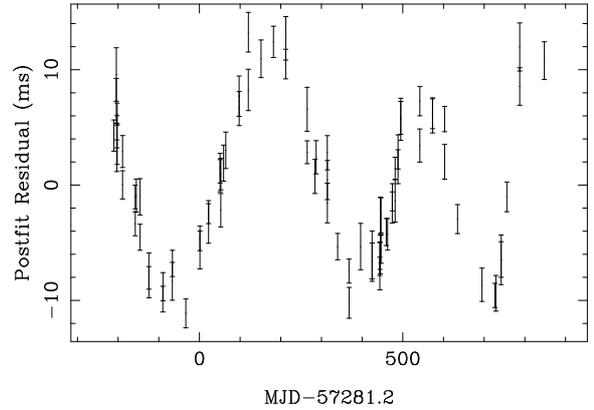}
\caption{Timing residuals after fitting out the cubic trend (timing solution) from the 
phase-shift measurements of the 1100-day data.}
\label{fig:bestfit}
\end{figure}

Before proceeding to analyze the 1100-day data, we confirmed the previous results 
(see Table.~\ref{tab:compare}) by actually fitting the 500-days data
with the same data used to infer $n=0.03$ \citep[25 {\it {Swift}} observations made
between MJDs~57070 and 57546;][]{Marshall2016}. We find $\nu=19.696323386(2)\rm \ s^{-1}$,
$\dot \nu=−2.528649(1) \times 10^{-10} \rm \ s^{-2}$, and $\ddot \nu=0.16(3)\times10^{-21}\rm \
s^{-3}$, yielding $n=0.05(1)$. These are consistent with the previous solution within 
statistical uncertainties.
We found no clear timing noise in the data as \citet{Marshall2016} already noted.
Thus, we used these 500-day data and the timing
solution of \citet{Marshall2016} as our starting points and gradually added new data
taken from MJDs 57546 to 58161 in the analysis, extending the baseline to 1100\,days.
Although the initial solution was valid for the first 500-day data, we found that the
phase shifts get larger with time and that the profiles do not phase-connect to the
initial one after $\sim$30 additional days.
We therefore refined the timing solution to phase-connect the later profiles
and repeated this to the last data set to construct a long-term 1100-day timing solution.
After fitting the rotational behavior (timing solution up to the cubic term),
there still appear to be relative phase shifts between the pulse profiles measured at different epochs (Fig.~\ref{fig:bestfit}). 
These timing residuals are the timing noise.
We also show the pulse profile constructed with the whole 1100-day data in the bottom panel
of Figure~\ref{fig:marshallpf} which is similar to the initial one (Fig.~\ref{fig:marshallpf} upper panel).
The new timing solution is presented in Table~\ref{tab:compare}.


We note that the timing noise pattern in Figure~\ref{fig:bestfit} is similar to that
measured by \citet{Livingstone} and \citet{Ferdman} using $>$10\,y data.
Additionally, notice that even in the first $\sim$500\,days, there is a residual
long-term trend (timing noise) which was not seen when using the previous timing solution.

\subsection{Timing noise\label{timingnoise}}

\begin{figure}
\centering
\includegraphics[width=80mm]{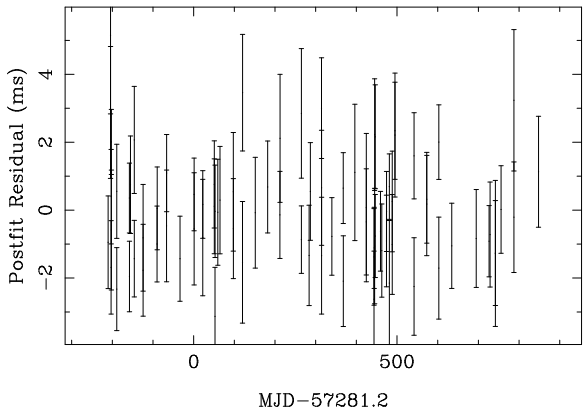}
\caption{Timing residuals obtained in {\tt TEMPO2} after whitening the timing noise.}
\label{fig:bestfit2}
\end{figure}

Physical origin and evolution of timing noise is not yet well understood and is therefore unpredictable.
Only by using a long baseline can timing noise (high-order polynomials) be 
distinguished from the timing solution (low-order polynomials). 
However, the concern is that there can be some mixing between the solution and the
noise. Thus, the timing solution may change a little if we model the timing noise
with high-order polynomials or sine functions and fit the data 
(e.g., Fig.~\ref{fig:bestfit}) to include the noise model.
A standard way to model and fit the timing noise is to use the power spectrum
of the timing residual (whitening). This method can be applied to the residuals
(Fig.~\ref{fig:bestfit}) with the {\tt TEMPO2} package
\citep{Hobbs}.\footnote{http://www.atnf.csiro.au/research/pulsar/tempo2/index.php?n\\=Main.T2psrparms}
We performed `whitening' using {\tt TEMPO2}, and the whitened residuals are shown in Figure~\ref{fig:bestfit2}.
The changes in the timing solution due to the whitening are small (insignificant).
As a crosscheck, we fitted the phase residual data with a timing noise model,
polynomials or sinusoidal functions and find that $n$ changes from $0.163$
(before whitening) to $0.164$ (after whitening). These two are statistically consistent.
We found that the change in the timing solution is very small in this case as well (see Table \ref{tab:compare}).

\section{Discussion\label{sec:discussion}}

We measured the timing properties of PSR~B0540$-$69 using the {\it Swift} data covering 1100\,days.
The timing solution obtained with the more robust $Z_1^2$ test is consistent with that measured
with the phase-coherent method. The measured $\nu$, $\dot\nu$, and $\ddot \nu$ with the phase-coherent
method imply a braking index of $n=0.163\pm0.001$. We further attempted to model or whiten the timing noise
and found that the timing solution is not sensitive to the timing noise models.

The solution we obtained for PSR B0540$-$69 is different from the previous ones \citep[see Table.~\ref{tab:compare};][]{Livingstone, Ferdman, Marshall2016}.
The discrepancy with those in \citet{Livingstone} or \citet{Ferdman} can be easily understood 
because our measurements were done far later in time from theirs; PSR B0540$-$69 is a young and 
active pulsar, and thus, the timing properties might have changed due to glitches or 
a spin-state transition (significant change in the rotational properties). 
Indeed \citet{Marshall2015} suggested a spin-state transition in MJD~55900 after
two glitches in MJD~51348 and MJD~52925; hence, solutions measured before \citep[e.g.,][]{Livingstone,Ferdman}
and after \citep[this work and ][]{Marshall2016} can differ.
The discrepancy with that of \citet{Marshall2016} is small but still statistically significant.
We attribute the discrepancy to the timing noise. Perhaps, their results may be biased
by undetermined timing noise. By properly isolating and accounting for the timing noise, we were able
to determine the timing solution more accurately.

Based on our new solution, we can estimate the bias in the previous result due to the timing noise.
As shown in Figure~\ref{fig:bestfit}, the timing noise
in the first $\sim$500\,days seems to follow a cubic trend with an amplitude of $\sim$10\,ms ($\phi_{\rm max}\approx0.2$).
We therefore modeled the residuals with a cubic function: $\phi(t) = at^3+bt^2+ct+d$,
where we reset the zero of $t$ to be at MJD 57546.08, the last data point in the 500-days time span.
We find $a=-3.82\times 10^{-22}\ \rm s^{-3}, b=1.73\times 10^{-15}\ \rm s^{-2}$, and 
$c=2.22\times 10^{-8}\ \rm s^{-1}$ in the fit.
These values can be compared with $\phi{(t)}$ in section~\ref{coherent}, resulting in
$\ddot \nu_{\rm bias}=6a=-2.3\times 10^{-21}\ \rm s^{-3}$,
$\dot \nu_{\rm bias}=2b=3.5\times 10^{-15}\ \rm s^{-2}$, and $\nu_{\rm bias}=c=2.22\times 10^{-8}\ \rm s^{-1}$,
which are sufficient for explaining the discrepancy. 
Hence, we argue that indeed the previous solution is biased due to the timing noise. 
This concern, however, was already noted in \citet{Marshall2016}; the authors
argued that the effects of the timing noise would not be large.
Our quantitative analysis suggests that the impact of the timing noise on the solution and $n$
is statistically significant but is not very large in absolute values as they argued.

\begin{table*}[ht]
\centering
\caption{Several timing solutions for PSR~B0540$-$69 obtained in previous works as well as this work\label{tab:compare}}
\scriptsize{
\begin{tabular}{lccccc}
\toprule
& Marshall & This Work (AW)$^{\rm 1}$ & This Work (BW)$^{\rm 1}$\\
\midrule
Observation date(MJD) & 57070$\sim$57546 & \multicolumn{2}{c}{57070$\sim$58068} \\
Epoch (MJD) & 57281.24 & \multicolumn{2}{c}{57281.24} \\
$\nu$ (Hz) & 19.6963233901(22) & 19.696323365(1) & 19.696323365(2)  \\
$\dot\nu$ ($10^{-10} s^{-2}$) & $-$2.5286507(15) & $-$2.528664(1) & $-$2.528664(2) \\
$\ddot\nu$ ($10^{-21} s^{-3}$) & 0.104(41) &  0.531(4) & 0.531(8) \\
Braking index (n) & 0.031(13)& 0.163(1) & 0.164(1)\\
\bottomrule
\end{tabular}}
\tabnote{$^{\rm 1}$ AW means After whitening, BW means before
whitening}
\end{table*}

As noted above, the most intriguing features of PSR B0540$-$69 are the
extremely small value and the large change of $n$ as suggested previously. We
arrive at the same conclusion with our study performed with a more accurate timing
solution; we confirmed the the main conclusion of \citet{Marshall2016}. Although such a small value of $n$ and large
change \citep[likely accompanied with the state change;][]{Marshall2015} are not
yet well understood, it is suggested that evolution of magnetic field in
the neutron star \citep{Gourgouliatos}, changes of mechanical properties
of the pulsar \citep{sedrakian}, and/or a putative accretion disk 
\citep[e.g., accretion rate;][]{Menou} may result in a small value of $n$ and the
change. We found that $n=0.163$ was larger than the previous one ($n=0.031$).
While this difference could be due to timing noise, it may also imply some
changes in the pulsar or in the putative disk.

Timing changes often occur contemporaneously with a radiative change but in PSR B0540$-$69
no significant evidence for a radiative change was found \citep{Marshall2015}.
However, PSR B0540$-$69 is in a crowded region of the sky and is surrounded by a PWN \citep{Kaaret}.
The 16$''$ angular resolution of {\it Swift} is not enough to resolve the pulsar from the 4$''$ nebular
or the nearby sources. Furthermore, the {\it Swift} observations are all in the timing mode without any 2D images.
Thus, unchanging (constant in time) contamination from the nearby sources and the PWN may dominate the flux,
making accurate measurements of small changes in the pulsar flux hard.
Future {\it Chandra} observations ($0.5''$ angular resolution) for spatially resolved spectroscopy \citep[e.g.,][]{An2014}
may be useful to measure a change in the flux of the pulsar and/or the nebula.

We finally note that a further timing analysis with longer ($>$500\,days) {\it Swift} data
has been performed by \citet{Marshall2018}.
In this AAS abstract, they reported that the braking index $n$ of PSR B0540$-$69 is still near 0, but details
on the timing solution are not avaliable. It would be interesting to check if they detected the timing
noise and if their timing solution agrees with ours. This can be done in the near future.

\acknowledgments

This research was supported by the Basic Science Research Program through
the National Research Foundation of Korea (NRF)
funded by the Ministry of Science, ICT \& Future Planning (NRF-2017R1C1B2004566).

\end{document}